\documentclass[onecolumn,preprintnumbers,aps,prl,floatfix,superscriptaddress]{revtex4}
\usepackage{epsfig}
\usepackage{graphics,graphicx,dcolumn,bm,fleqn,epic,eepic,float}
\usepackage{amssymb,amsmath,multirow,rotate,color,float}
\usepackage{times}
\usepackage{verbatim}
\usepackage{subfigure}
\usepackage{color}

\definecolor{red}{rgb}{1,0,0}

\definecolor{blue}{rgb}{0,0,1}

\begin{document}
\title{Emergence of scaling in human-interest dynamics}

\date{\today}

\author{Zhi-Dan Zhao}
\affiliation{Web Sciences Center, University of Electronic Science and
Technology of China, Chengdu 610054, China.}
\affiliation{School of Electrical, Computer and Energy Engineering, Arizona
State University, Tempe, Arizona 85287, USA}

\author{Zimo Yang}
\affiliation{Web Sciences Center, University of Electronic Science and
Technology of China, Chengdu 610054, China.}

\author{Zike Zhang}
\affiliation{Web Sciences Center, University of Electronic Science and
Technology of China, Chengdu 610054, China.}
\affiliation{Institute for Information Economy, Hangzhou Normal University, Hangzhou 310036,
China}

\author{Tao Zhou}
\affiliation{Web Sciences Center, University of Electronic Science and
Technology of China, Chengdu 610054, China.}

\author{Zi-Gang Huang}
\affiliation{Institute of Computational Physics and Complex Systems, Lanzhou University,
Lanzhou, Gansu 730000, China}
\affiliation{School of Electrical, Computer and Energy Engineering, Arizona
State University, Tempe, Arizona 85287, USA}

\author{Ying-Cheng Lai}
\affiliation{School of Electrical, Computer and Energy Engineering, Arizona
State University, Tempe, Arizona 85287, USA}

\maketitle

{\bf Human behaviors are often driven by {\em human interests}. Despite
intense recent efforts in exploring the dynamics of human behaviors,
little is known about human-interest dynamics, partly due to the extreme
difficulty in accessing the human mind from observations. However, the
availability of large-scale data, such as those from e-commerce and
smart-phone communications, makes it possible to probe into and
quantify the dynamics of human interest. Using three prototypical
``big data'' sets, we investigate the scaling behaviors associated
with human-interest dynamics. In particular, from the data sets
we uncover power-law scaling associated with the three basic
quantities: (1) the length of continuous interest, (2) the return time
of visiting certain interest, and (3) interest ranking and transition.
We argue that there are three basic ingredients underlying human-interest
dynamics: \emph{preferential return to previously visited interests,
inertial effect, and exploration of new interests}. We develop a biased
random-walk model, incorporating the three ingredients, to account for
the observed power-law scaling relations. Our study represents
the first attempt to understand the dynamical processes underlying human
interest, which has significant applications in science and engineering,
commerce, as well as defense, in terms of specific tasks such as
recommendation and human-behavior prediction.}

A fundamental feature of a human society is that its individuals
possess all kinds of interests, the driving force of many human
behaviors. Some interests may
last for a lifetime while others can fade away in short time. From
time to time our interests also change. In the modern society that
we live in, all kinds of attractions and temptations emerge and
disappear on a daily basis. Does this mean that the evolution of
our interest is mostly random? Or are there intrinsic dynamical
rules that govern how human interests evolve with time? To answer
these questions was deemed to be extremely difficult, due to the
lack of appropriate means to characterize human mind and to measure
quantitatively how it changes with time. Yet the questions are
fundamental in science, and any revelation of the dynamics of
human interest may have significant applications in commerce,
medical sciences, and even defense. In particular, in commerce,
adequate knowledge of customer interests and how they change with time
are key to the success of many businesses as such knowledge can
be of tremendous value to advertisement design and product promotion.
In psychiatry, a good understanding of patients' interests may help
generate accurate diagnosis and devise effective therapeutic
approaches. In defense, timely and reliable assessment of certain group or
individuals' interests and their time evolution can help predict
the group or individuals' possible future behaviors and actions.
Apparently, all these rely on human-interest dynamics' being not
completely random.

There have been efforts in modeling and understanding human behaviors
that are essential to many social and economical phenomena, with
significant applications in areas ranging from resource allocation
and transportation control to epidemic prediction and personal
recommendation~\cite{Barabasi2007,Castellano2009,guimera2012,Lu2012}. The
pursuit has been facilitated greatly by the advances in information
technology, especially by the availability of massive Internet data
and resources~\cite{lazer2009}. However, to probe into human-interest
dynamics is more challenging, due to the difficulty in characterizing
human interests and traditional lack of data sets from which the
underlying dynamical processes may be deduced. In recent years ``big data''
sets, such as those from e-commerce or mobile-phone communications,
become commonly available, making it possible to quantify human
interests and to infer their intrinsic dynamics. As a branch of the
science of ``Big Data,'' the field of human-interest dynamics
is at its infancy.

A viable approach to probing into human-interest dynamics is to use
data analysis as a getaway to uncover various phenomena and possible
scaling laws. Guided by this principle, in this paper we explore two
e-commerce data sets (Douban, Taobao) and one communication data set
[Mobile-Phone Reading (MPR)], and focus on three
issues: statistical distribution of the time that an interest lasts,
distribution of the return time to revisit a particular interest, and
interest ranking and transition. Considering the large number of factors
that can affect human interest, such as the specific activity contents and
distractions of the individual's attention, it seems plausible
that the underlying dynamics be completely
random~\cite{wu2007,weng2012,ye2012}. Indeed, a widely used assumption
is that of the Markovian type of dynamics for individuals' online
behaviors, in which an online user's next action depends not on
his/her history of interests but on the current interest
only~\cite{Brin1998,craswell2007,fagin2001}. However, there is
recent evidence~\cite{Meiss2010,Chierichetti2012}
of deviations from the Markovian dynamics. Our systematic analysis
of the three data sets reveals an unequivocal signature of the
power-law scaling behavior characteristic
of non-equilibrium complex systems and, consequently,
indicates the existence of intrinsic dynamical rules governing the
human-interest dynamics. Based on the empirical analysis, we identify
three basic ingredients underlying the dynamics: preferential
return, inertia effect and exploration. A mathematical model
incorporating these ingredients is then developed to account for the
observed power-law scaling behaviors. Our study represents the first
systematic attempt to probe into the dynamics of human interest, and
we expect our finding and model to have broad applications.

We note that, in the study of human behaviors, heavy-tailed type of
statistical features, e.g., those in the inter-event time
distributions~\cite{Barabasi2005,Oliveira2005,Dezso2006,Zhou2008,Goncalves2008,Wu2010},
have been uncovered recently. Such a non-Poisson type of distribution
implies, e.g., that the bursts of rapidly occurring events are typically
separated by long periods of inactivity. Various mechanisms have been
proposed to explain the heavy-tailed inter-event statistics, such as
the highest-priority-first queue model~\cite{Barabasi2005,Vazquez2006},
Poisson probability model~\cite{Malmgren2008,Malmgren2009}, varying
interest~\cite{Han2008}, memory effects~\cite{Vazquez2007}, and human
interactions~\cite{Oliveira2009,Min2009,Wu2010}. Non-Poisson,
heavy-tailed type of statistics also arise in human mobility
trajectories~\cite{Brockmann2006,Gonzalez2008,Rhee2011}, and
mathematical models have been proposed to account for the non-Markovian
type of dynamics underlying the human mobility, such as those based on
exploration and preferential return~\cite{Song2010}, hierarchy of
traffic systems~\cite{Han2011}, and regular mobility~\cite{Yan2011}.
Variances in the statistical behaviors of human mobility were also
reported~\cite{Huberman1998,stehle2010,Karsai2012}. The distinct
feature of our work is its focus on human-interest dynamics.

\section{Results} \label{sec:result}

We analyze three massive data sets: two from e-commerce, namely,
{\em Douban} and {\em Taobao}, and one from mobile-communication,
i.e., {\em MPR}. We focus on the scaling of three quantities:
(1) the time interval $l$ that an individual stays within the same
interest, defined as the length of a sequence of clicks within the same interest
category (defined in Methods), (2) the time interval $\tau$ that an individual
returns to visit the same interest category, defined as the sequence of clicks
between two visits to the same interest, representing a kind of {\em memory}
effect in the dynamics of interest, and (3) the frequencies of visit of an
individual to different interests, which can be used to rank this individual's
particular interests.

\begin{figure*}
\begin{center}
\epsfig{figure=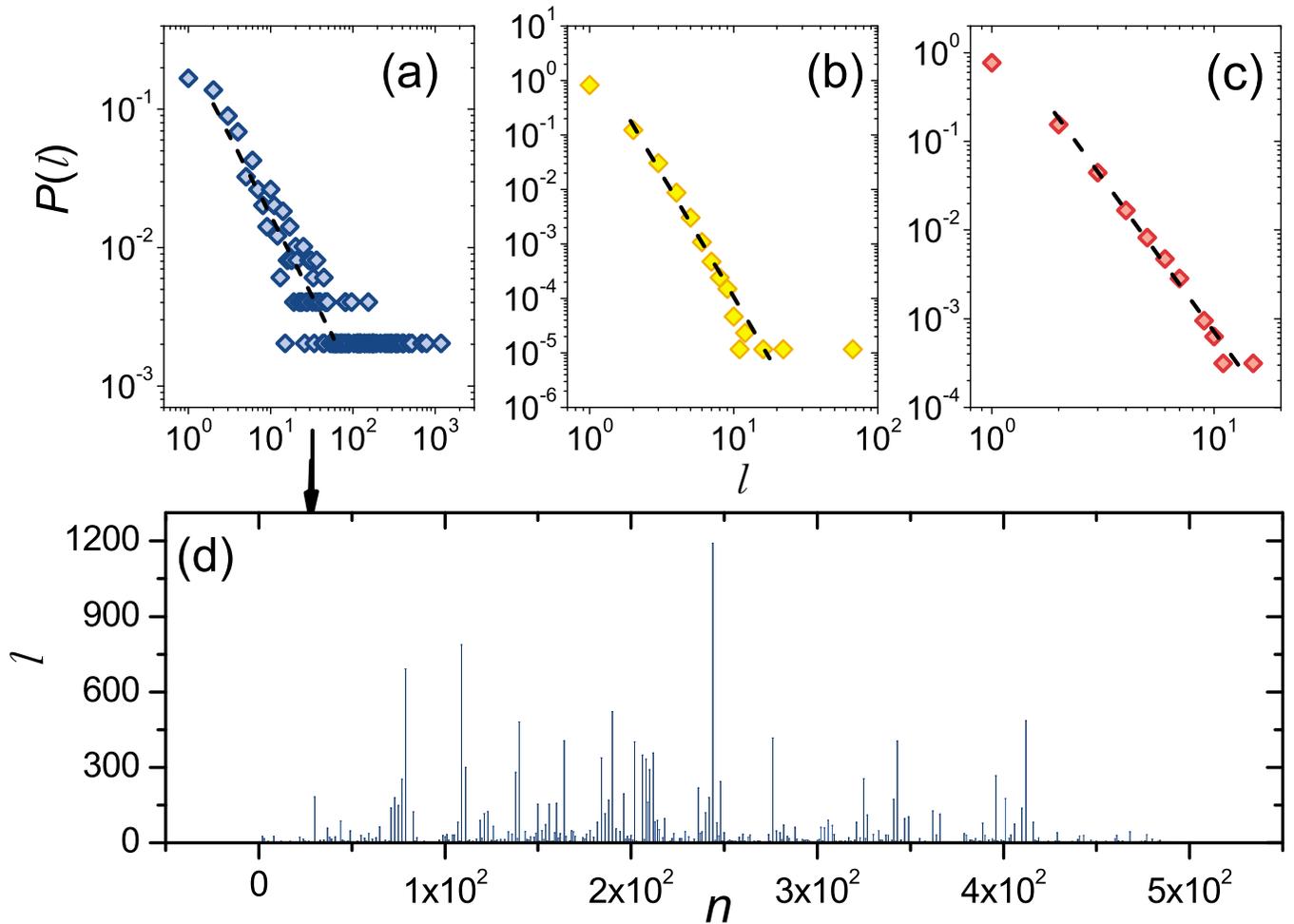,width=\linewidth}
\end{center}
\caption{
\baselineskip=12pt
{\bf Distribution of interest-dwelling time}.
(a-c) Probability distributions $P(l)$ of the time interval $l$ of consecutive
visits to the same interest for three representative individuals, each from one
of the three data sets ({\em Douban}, {\em Taobao}, and {\em MPR}),
where the numbers of interests are 3, 24, and 44, respectively.
The numbers of clicks ($N_a$) for the three cases are are $18396$,
$106571$, and $4398$, respectively. The three distributions can be
fitted as power-law $P(l) \sim l^{-\alpha}$, with exponents
$\alpha \approx 1.16, 4.02$ and $3.35$, respectively (the values of
the exponent $\alpha$ are estimated using the maximum-likelihood
criterion~\cite{Clauset2009}). Panel (d) shows the various values of
$l$ as they appear with time, where $n$ is the event index (an integer variable).}
\label{fig:P_l}
\end{figure*}

{\em Power-law scaling of interest interval $l$.} A number
of approaches have been proposed to characterize an individual's interests, such
as the interest profile~\cite{Lam2001}, contextual
information~\cite{White2009}, distinct visited subpages~\cite{Chmiel2009},
and service items~\cite{Yang2011}. Taking advantage of the nature of our large data
sets, we use categories to characterize an individual's interests, which can be,
for example, music, books and movies on {\em Douban}, clothing, footwear, and
toys in {\em Taobao}, love stories and science fictions on {\em MPR}, and so on.
Figure~\ref{fig:P_l}(a) shows, for a typical individual on {\em Douban}, the
distribution $P(l)$ of $l$ visiting different interest categories, which
exhibits a power-law scaling: $P(l) \sim l^{-\alpha}$. The long tail associated
with the power-law scaling indicates that the individual tends to spend an
abnormally long time visiting certain interests during browsing. Similar scaling
behaviors have been found for users on {\em Taobao} and {\em MPR}, as shown in
Figs.~\ref{fig:P_l}(b) and \ref{fig:P_l}(c), respectively.
A typical sequence that the values of $l$
corresponding to an identical interest appear is shown in Fig.~\ref{fig:P_l}(d).
From Fig.~\ref{fig:P_l}(d), we observe a highly non-uniform
behavior in the values of $l$, which gives rise to the power-law distribution
in Fig.~\ref{fig:P_l}(a). We have examined many individuals from the three
data sets, and found similar power-law behaviors. In fact, the distribution
of $l$ for {\em all} users from any particular data set exhibits a robust
power-law scaling (Fig.~S1 in Supplementary Information). The universal power-law
scaling observed for all cases implies substantial derivation of the
human-interest dynamics from that of the Markovian process because, for such
a fully random process, the scaling of $l$ would be exponential~\cite{Kingman1963}.

\begin{figure*}
\begin{center}
\epsfig{figure=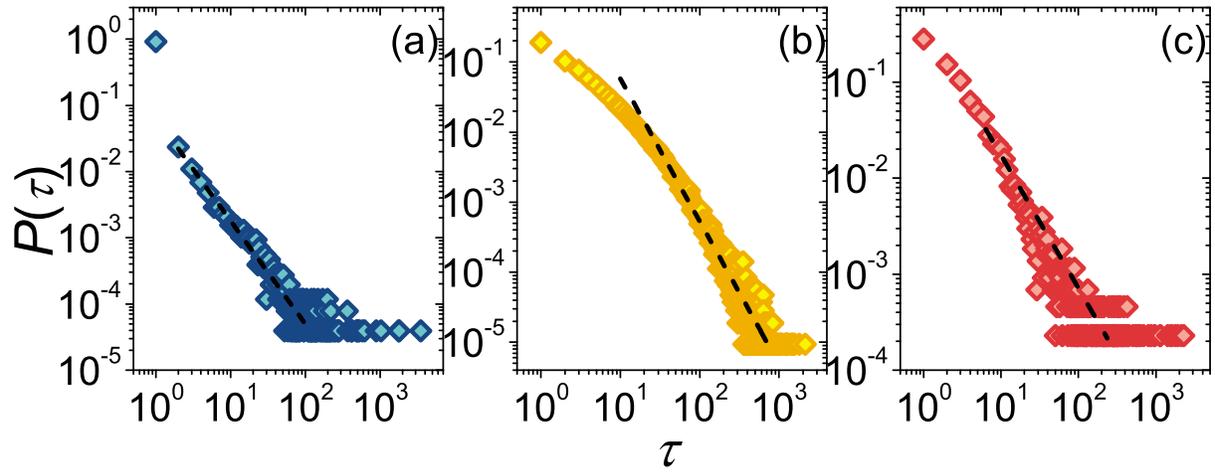,width=\linewidth}
\end{center}
\caption{\baselineskip=12pt
{\bf Memory effect of human interest dynamics.} (a-c) For the data sets in
Figs.~\ref{fig:P_l}(a-c), respectively, power-law distributions ($\tau^{-\beta}$)
of the time $\tau$ taken to revisit the same interest. The values of the fitted
exponent $\beta$ are approximately $1.58$, $2.04$, and $1.41$ for (a-c), respectively.}
\label{fig:P_n}
\end{figure*}

{\em Memory effect in human-interest dynamics.} Memory, as one of the key
attributes of human being, has been widely studied in the
past~\cite{Vazquez2007,Yamasaki2005,Goh2008,Cai2009,Han2008,Szell2012,Karsai2012}.
We observe from our data sets that, often, an individual tends to return to
specific interests that he/she has recently visited with relatively higher
probabilities than those visited long ago. For example, even when an interest
had been visited many times in the past, if the most recent visit dates back
one year or longer, the probability of revisiting is lower as compared with
that associated with another interest that was visited
merely a week ago. But would the probability
that an interest is revisited after a very long time be exponentially small?
To answer this question, we calculate the distribution of the return
time~\cite{Szell2012} $\tau$, the time interval that an individual revisits the
same interest after the last visit. Typical distributions from three individuals,
one from each data base, are shown in Figs.~\ref{fig:P_n}(a-c), which can again
be well fitted by power laws: $P(\tau) \sim \tau^{- \beta}$, with the exponent
$\beta$. While $P(\tau)$ is higher for small values of $\tau$, the probability of
the occurrence of very large values of $\tau$ is, surprisingly, not exponentially
small, indicating that such events can indeed occur. An important implication
is that, both short-term and long-term memories can shape the human-interest
dynamics. Similar results are obtained for many other users (Fig.~S3 in
Supplementary Information). Additionally, the distribution of $\tau$ for
{\em all} users from any particular data set exhibits a robust power-law
scaling (Fig.~S1 in Supplementary Information).

{\em Interest ranking and transition among interests}. An individual can
possess a number of interests, which can be ranked in terms of the respective
frequencies of visit. In a given (large) time interval,
an individual can focus on different interests, giving rise to a kind of
``transition'' among the interests. The interest ranking and transition
are important not only for the study of human dynamics~\cite{Barabasi2005,Song2010}
and decision-making~\cite{Busemeyer1993,salganik2006},
but also for applications such as behavior prediction and search-algorithm design.

\begin{figure*}
\begin{center}
\epsfig{figure=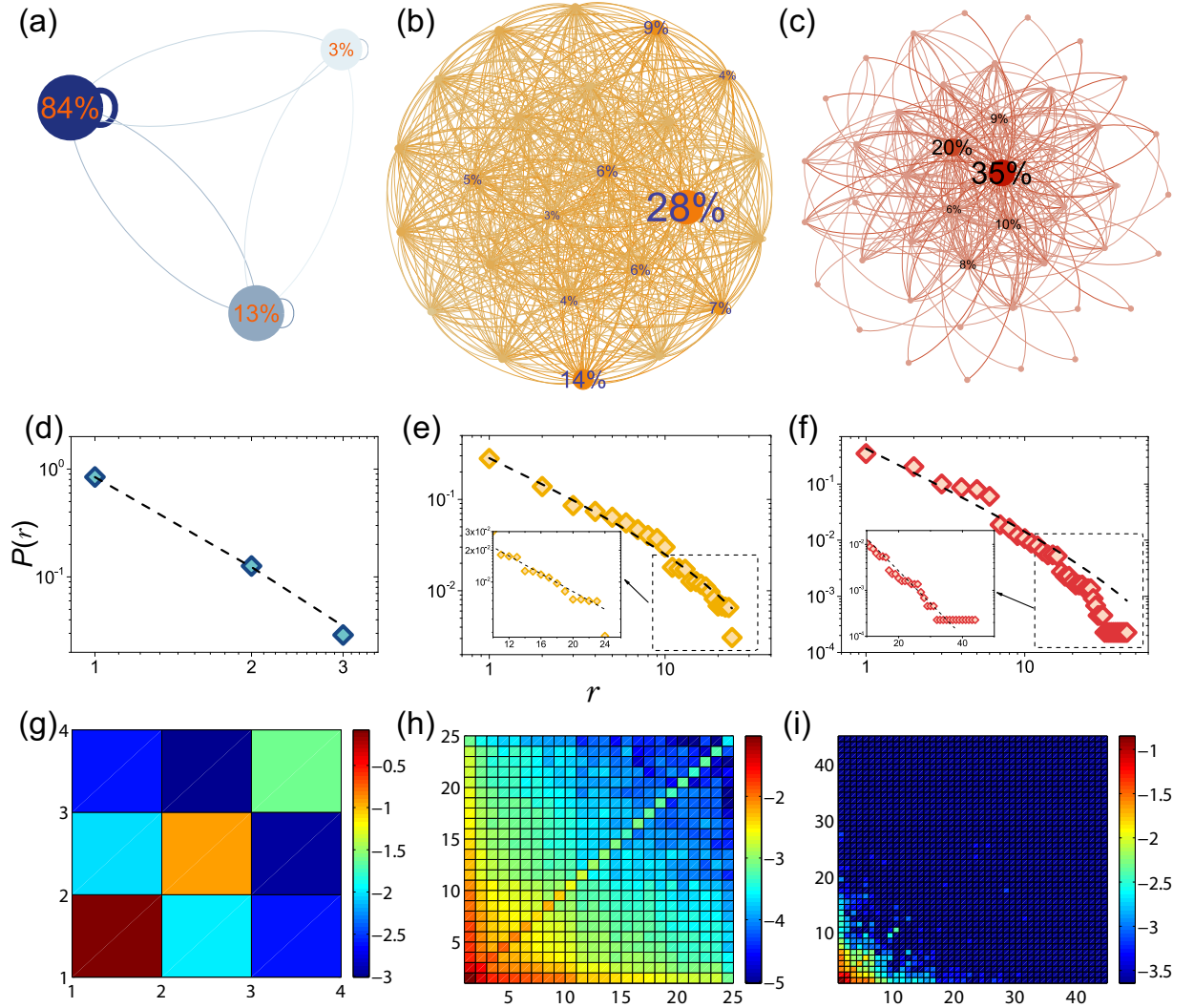,width=\linewidth}
\end{center}
\caption{\baselineskip=12pt
{\bf Interest-transition network and transition probabilities}.
(a-c) For the three individuals represented in Figs.~\ref{fig:P_l}(a-c), the
respective transition networks, where nodes correspond to distinct interests,
a self loop represents the dwelling time in the same interest category, and
the weighted links characterize the interest transitions. A few highly
frequently visited interests are marked. (d-f) Truncated power-law scaling in
the rank distribution: $P(r)\propto {r^{-\gamma}}\exp{( - r/{\rm S})}$,
where the fitted values of the exponent $\gamma$ and the numbers of interests
are $(\gamma,S)=(2.40,3)$ ({\em Douban}), $(\gamma,S)=(0.89,24)$ ({\em Taobao}),
and $(\gamma,S)=(1.39,44)$ ({\em MPR}). (g-i) Two-dimensional representation
of the interest-transition probabilities for the three networks in (a-c),
respectively.}
\label{fig:transition}
\end{figure*}

A convenient way to assess the interest-transition pattern for an individual
is to use a network representation, where nodes denote
different interests with sizes determined by their ranks, links
correspond to the observed transitions among the interests, and the dwelling time
in any particular interest is represented by a self loop. Such network
representations have also been used in other contexts such as transportation
dynamics~\cite{Banavar1999}, citations~\cite{Rosvall2008}, and human-mobility
behaviors~\cite{Song2010a}. Figures~\ref{fig:transition}(a-c) show examples
of the transition networks of one typical individual from each of the
three data sets, respectively. Setting the most frequently visited interest
to have rank $r = 1$ and the successively less frequently visited interests
to have ranks $r = 2$, 3, and so on, we can generate a distribution of the
interest rank for each individual, examples of which are shown in
Figs.~\ref{fig:transition}(d-f). In all cases, such a rank distribution can
be approximately fitted by the following exponentially truncated
power-law: $P(r) = r^{-\gamma}\exp{(-r/S)}$, where $\rm S$ is the number
of distinct interests that the individual has selected. Note that this
truncated power-law is with respect to an individual. When the collective behavior
of a large number of individuals is considered, the signature of the
exponential truncation diminishes and the scaling of $P(r)$ can be
better fitted by a power law (see Fig.~S1 in Supplementary Information).
This is similar to the power-law ranking distribution observed in the
collective human-mobility patterns~\cite{Gonzalez2008,Song2010,Bagrow2012}
where the distribution is with respect to the actual locations that the
individual visits physically.

{\em Model of human-interest dynamics}.
To gain insights into the development of a quantitative model describing
the dynamics of human interest, we investigate the transition probability
defined as $p(i,j) = n(i,j)/\sum_{i,j}n(i,j)$, where $n(i,j)$ is the number
of switchings from interest $i$ to $j$. Examples of the transition
probabilities, those corresponding to the respective transition networks
in Figs.~\ref{fig:transition}(a-c), are shown in Figs.~\ref{fig:transition}(g-i)
in the two-dimensional representation of $i$ and $j$. We observe two key
features: (i) $p(i,j)$ exhibits relatively large values for transitions
among the highly ranked interests (note that $r = 1$ corresponds to the
highest ranked interest), and (ii) the diagonal elements $p(i,i)$ have
relatively large values as well. The first feature suggests a kind of preferential
selection~\cite{Barabasi1999,Gonalves2009,Song2010,Meiss2010,Sorribes2011}
of interests: individuals tend to return to highly
ranked interests with relatively larger probabilities and stay in
these interests. The second feature indicates an inertial effect:
an individual tends to stay in the interest that he/she has already
been exploring. These two ingredients, preferential return and
inertia, plus an individual's desire to explore new interest, are
the basic ingredients underlying the human-interest dynamics, based
on which a phenomenological model can be developed.

\begin{figure*}
\begin{center}
\epsfig{figure=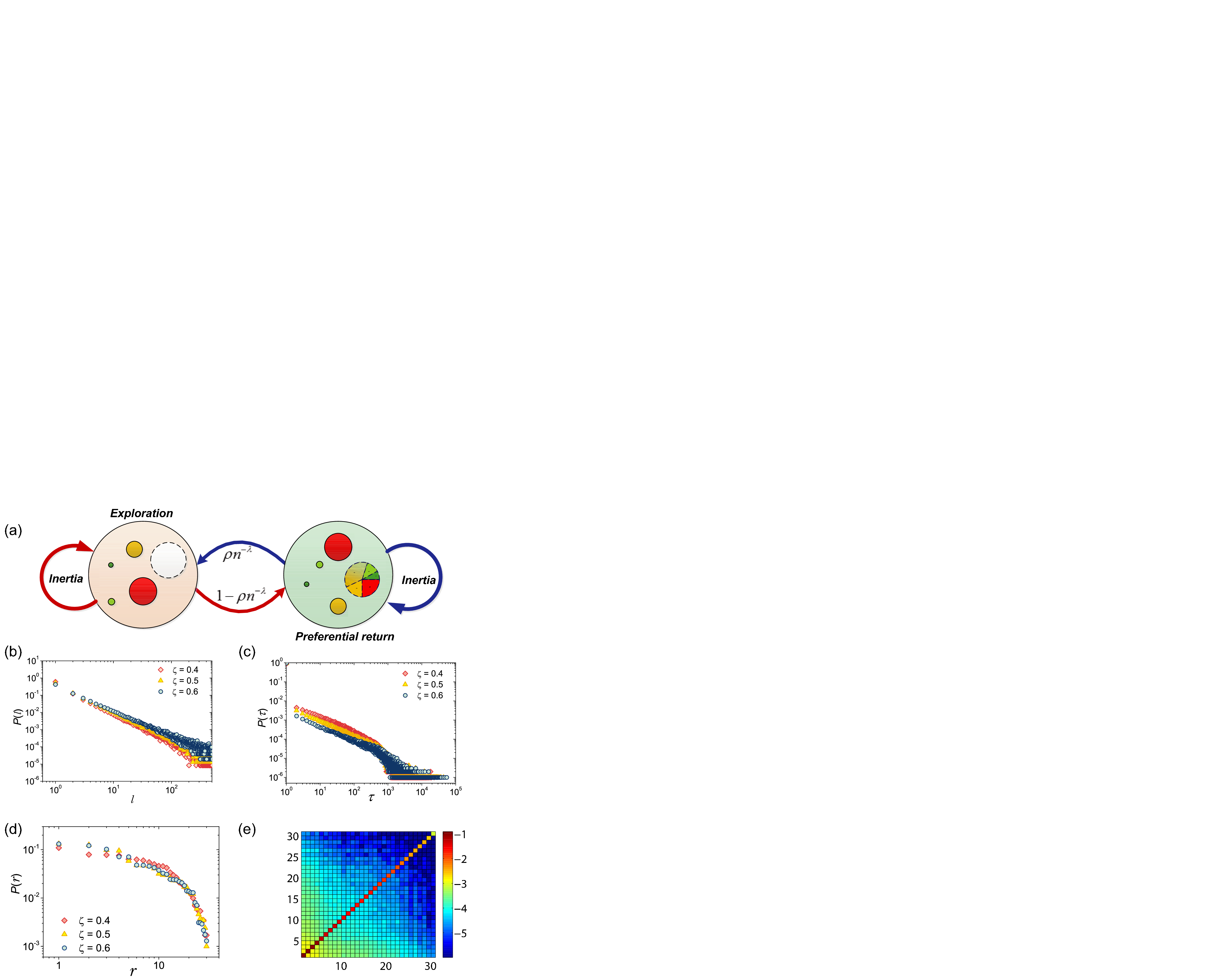,width=\linewidth}
\end{center}
\caption{
\baselineskip=12pt
{\bf Proposed model of human-interest dynamics and predicted scaling
relations}. (a) Schematic illustration of the model, where an individual
can enter one of the two dynamically complementary states at each hopping
step: exploring new interests with the probability $\rho n^{-\lambda}$ (the
state of ``Exploration'') or returning preferentially to a previously
explored interest with the probability $1 - \rho n^{-\lambda}$ (the state
of ``Preferential return''). Regardless of which state takes place,
one interest is selected and the inertial effect sets in, a microscopic process
that can be modeled as an excited random walk (ERW)~\cite{Antal2005}.
(b,c) Power-law scaling of $P(l)$ and $P(\tau)$, respectively.
(d,e) Predicted interest-ranking distribution and transition-probability
pattern, respectively. These results are obtained from model simulations
using $10^6$ time steps for the parameter setting of $\lambda = 0.4$ and
$\rho = 0.6$. For $P(l)$, analytic result can be derived:
$P(l) \sim l^{-(2-\zeta)}$, where $\zeta$ and $1-\zeta$ are the probabilities
of moving towards the ``right'' or the ``left,'' respectively. In (b-d),
three values of $\zeta$ are used: $\zeta = 0.4$, $\zeta = 0.5$, and
$\zeta = 0.6$. In (e), the value of $\zeta$ is $0.5$.}
\label{fig:model}
\end{figure*}

A schematic illustration of our model is shown in Fig.~\ref{fig:model}(a).
To initiate the dynamical evolution of interest, an individual has two
options: exploration of new interest or return to one of the previously
visited interests, with probability $\rho n^{-\lambda}$ and
$1 - \rho n^{-\lambda}$, respectively, and $0 < \rho \le 1$ and $\lambda > 0$
are parameters~\cite{Song2010,Szell2012}, where $n$ denotes the number of
hopping-events among different interests, which is obtained by merging
the same interest in click-event series into one. For example, the
click-event series {1, 1, 2, 2, 2, 1, 3} with $7$ actions can be
transformed into the following hopping-event series: {1, 2, 1, 3}, where $n = 4$.
In the exploration state, individual visits a new interest and continuously
browses the same interest, due to the effect of inertia. At a ``microscopic''
level, inertial browsing can be regarded as an excited random-walk process
(ERW)~\cite{Antal2005}. If the individual returns to a set of previous
revisited interests, he/she preferentially selects an interest category
to browse according to the prior probability of visit to the same interest.
Once a particular interest is chosen, the inertial effect sets in and the
individual has the tendency to stay in the same interest category. The microscopic
browsing behavior again can be modeled by an excited random-walk process.
A detailed mathematical analysis of the model in Fig.~\ref{fig:model}(a)
can be found in Supplementary Information. Examples of the predicted
scaling relations are illustrated in Figs.~\ref{fig:model}(b-d) (with more
examples in Supplementary Information), which are consistent
with those uncovered from real data as exemplified in
Figs.~\ref{fig:P_l}-\ref{fig:transition}.

\section{Discussion}

Despite recent efforts in human-mobility
dynamics~\cite{Barabasi2005,Oliveira2005,Dezso2006,Zhou2008,Goncalves2008,Wu2010},
little is known about human-interest dynamics. We aim to explore the fundamental
mechanisms underpinning the human-interest dynamics through a completely
data-driven approach. In particular, we have analyzed three large-scale
data sets: two from e-commerce and one from mobile communication, and
uncovered the emergence of scaling behaviors in a number of fundamental
quantities. These are the interval $l$ to stay in an interest,
the time interval $\tau$ to return to a previously visited interest, and
the interest-ranking distribution. A detailed analysis of the patterns of
the transition probabilities among different interests suggests preferential
return, inertia, and exploration as the three basic dynamical ingredients
underlying the human-interest dynamics, enabling us to construct a
phenomenological, random-walk based model. The model captures the essential
features of the human-interest dynamics in that it is constructed based on
generic ingredients extracted from real data, and it is capable of
reproducing the scaling laws observed from data. The model, however,
may still be idealized as it cannot predict the scaling exponents.
To develop a more predictive model, additional effects must be included, such as
individual's memory effect~\cite{Vazquez2007,Meiss2010,Karsai2012}, cognitive
activities~\cite{Busemeyer1993,Sorribes2011}, and the specific
web categories, etc. Nonetheless, the current model provides a phenomenological
framework where the basic properties and scaling behaviors associated with
human-interest dynamics can be explained.

The scaling laws uncovered from data and the dynamical model developed
accordingly can be applied to addressing significant problems ranging
from human-behavior prediction and the design of search
algorithms~\cite{Vespignani2009,Song2010} to controlling spreading
dynamics~\cite{Balcan2011,Zhao2012}. As a demonstration, we have
quantified the degree of predictability of user-behavior patterns underlying
the three data sets by using the statistical measures of entropy and
Fano inequality~\cite{Song2010}, with the result that such patterns are
in fact quite predictable, despite the apparent randomness in the
human-interest dynamics (see Supplementary Information).

\section{Methods} \label{sec:method}

{\em Data collection}.
The massive data sets used in this article are from large-scale real
e-commerce and communication systems: {\em Douban}, {\em Taobao}, and
{\em MPR}. For fair comparison, in each data set we focus on users who
performed at least 100 actions. Data description and basic statistical
properties are listed in Table~\ref{tab:originalpreference}.

(i) \emph{Douban}. The experimental data set is randomly sampled from Douban,
a major e-commerce company in China. It is similar to the Social Networking
Services (SNS) that allows registered users to record information and create
contents related to movies, books, and music, etc., yet it can also make
personalized recommendations for the registered users. In this data set, we
select 21,148 individuals, each executing at least 100 rating actions, from
which we can find historical information about the users, such as
user ID, item ID, rate, timestamps, and item types (considered as interest
types), etc. The sampling time resolution is one second.

(ii) \emph{Taobao}. The Chinese web site Taobao is one of the world's largest
electronic marketplaces. The browsing behaviors of users on Taobao are
recorded, and any user can browse and trade with any other users. Our data is
composed of all browsing behaviors of 34,330 users, each browsing more than
100 items in the time span between September 1 and October 28, 2011. For each
user, information is available such as the user ID, item ID, item
classes (regarded as interest types), timestamps, etc. The sampling time
resolution is one second.

(iii) \emph{MPR} - a widely used electronic reading tool. The usage of such a
mobile service reflects well customers' interests. We collected the reading
records of 19,067 users, each performing more than 100 reading tasks between
October 1 and October 31, 2011. The categories of books that each reader
chose are regarded as interests. The sampling time resolution is one day.

\begin{table}
\caption{Basic parameters of the three massive data sets studied in this
paper.}
\begin{tabular}{cccc}
\hline \hline
       Data Sets & \#Users & \#Time-span & Origins \\
\hline
       {\em Douban}  & 21,148 & 18 months & This article \\
       {\em Taobao}  & 34,330 & 2 months  & This article \\
       {\em MPR}     & 19,067 & 1 month   & This article \\
\hline \hline
\end{tabular}
\label{tab:originalpreference}
\end{table}

{\em Definition of length of interest interval $l$.}
Previous studies defined session as a sequence of Web pages viewed by a
user within a given time window, which has been widely used in modeling
and tracking individuals' navigation
behaviors~\cite{Spiliopoulou2003,Borges2007,Meiss2009,Fortunato2006,Gonalves2009}.
However, for characterizing human interest, this definition of session has
two deficiencies: (1) difficulty to split an individual's click sequence
into sessions~\cite{Meiss2009} due to the continuous nature of the
user online activities~\cite{Song2010,Zhou2012}, and (2) limit in the
data sets, due to the time resolution of MPR (day). Thus, we define the
interest duration $l$ as the length of a sequence of clicks within the
same interest category.

\vspace{0.75in}

\textbf{Acknowledgments}

We thank Dr Jumming Huang for the Douban data.
This work was supported by the NNSF of China (Grants No.11275003, 11222543),
and by the Fundamental Research Funds for the
Central Universities (Grant No. ZYGX2011YB024). YCL was supported by
AFOSR under Grant No.~FA9550-10-1-0083 and by NSF under Grant
No.~CDI-1026710.

\vspace{0.5in}
{\bf Supplementary Information} is linked to the online version  of
the paper at XXX.

\vspace{0.5in}
\textbf{Author Contributions.}
All authors participated in the design of the project. ZDZ and ZZ
performed data preparation and analysis. ZDZ, ZGH, ZY and TZ discussed the
model. ZGH and ZDZ did theoretical analysis. ZDZ and YCL wrote the paper.

\vspace{0.5in}

{\bf Author Information} Correspondence and requests
for materials should be addressed to YCL
(e-mail: Ying-Cheng.Lai@asu.edu)

\newpage


\begin{thebibliography}{10}
\expandafter\ifx\csname url\endcsname\relax
  \def\url#1{\texttt{#1}}\fi
\expandafter\ifx\csname urlprefix\endcsname\relax\def\urlprefix{URL }\fi
\providecommand{\bibinfo}[2]{#2}
\providecommand{\eprint}[2][]{\url{#2}}

\bibitem{Barabasi2007}
\bibinfo{author}{Barab{\'a}si, A.-L.}
\newblock \bibinfo{title}{The architecture of complexity}.
\newblock \emph{\bibinfo{journal}{IEEE Contr. Syst. Mag.}}
  \textbf{\bibinfo{volume}{27}}, \bibinfo{pages}{33--42}
  (\bibinfo{year}{2007}).

\bibitem{Castellano2009}
\bibinfo{author}{Castellano, C.}, \bibinfo{author}{Fortunato, S.} \&
  \bibinfo{author}{Loreto, V.}
\newblock \bibinfo{title}{Statistical physics of social dynamics}.
\newblock \emph{\bibinfo{journal}{Rev. Mod. Phys.}}
  \textbf{\bibinfo{volume}{81}}, \bibinfo{pages}{591--646}
  (\bibinfo{year}{2009}).

\bibitem{guimera2012}
\bibinfo{author}{Guimer{\`a}, R.}, \bibinfo{author}{Llorente, A.},
  \bibinfo{author}{Moro, E.} \& \bibinfo{author}{Sales-Pardo, M.}
\newblock \bibinfo{title}{Predicting human preferences using the block
  structure of complex social networks}.
\newblock \emph{\bibinfo{journal}{PLoS ONE}} \textbf{\bibinfo{volume}{7}},
  \bibinfo{pages}{e44620} (\bibinfo{year}{2012}).

\bibitem{Lu2012}
\bibinfo{author}{L{\"u}, L.} \emph{et~al.}
\newblock \bibinfo{title}{Recommender systems}.
\newblock \emph{\bibinfo{journal}{Phys. Rep.}} \textbf{\bibinfo{volume}{519}},
  \bibinfo{pages}{1--49} (\bibinfo{year}{2012}).

\bibitem{lazer2009}
\bibinfo{author}{Lazer, D.} \emph{et~al.}
\newblock \bibinfo{title}{Life in the network: the coming age of computational
  social science}.
\newblock \emph{\bibinfo{journal}{Science}} \textbf{\bibinfo{volume}{323}},
  \bibinfo{pages}{721} (\bibinfo{year}{2009}).

\bibitem{wu2007}
\bibinfo{author}{Wu, F.} \& \bibinfo{author}{Huberman, B.~A.}
\newblock \bibinfo{title}{Novelty and collective attention}.
\newblock \emph{\bibinfo{journal}{Proc. Natl. Acad. Sci. U. S. A.}}
  \textbf{\bibinfo{volume}{104}}, \bibinfo{pages}{17599--17601}
  (\bibinfo{year}{2007}).

\bibitem{weng2012}
\bibinfo{author}{Weng, L.}, \bibinfo{author}{Flammini, A.},
  \bibinfo{author}{Vespignani, A.} \& \bibinfo{author}{Menczer, F.}
\newblock \bibinfo{title}{Competition among memes in a world with limited
  attention}.
\newblock \emph{\bibinfo{journal}{Scientific Reports}}
  \textbf{\bibinfo{volume}{2}} (\bibinfo{year}{2012}).

\bibitem{ye2012}
\bibinfo{author}{Ye, M.}, \bibinfo{author}{Sandholm, T.},
  \bibinfo{author}{Wang, C.}, \bibinfo{author}{Aperjis, C.} \&
  \bibinfo{author}{Huberman, B.~A.}
\newblock \bibinfo{title}{Collective attention and the dynamics of group
  deals}.
\newblock In \emph{\bibinfo{booktitle}{Proc. 21st Int. Conf. WWW}},
  \bibinfo{pages}{1205--1212} (\bibinfo{organization}{ACM},
  \bibinfo{year}{2012}).

\bibitem{Brin1998}
\bibinfo{author}{Brin, S.} \& \bibinfo{author}{Page, L.}
\newblock \bibinfo{title}{The anatomy of a large-scale hypertextual web search
  engine}.
\newblock \emph{\bibinfo{journal}{Comp. Net. ISDN}}
  \textbf{\bibinfo{volume}{30}}, \bibinfo{pages}{107--117}
  (\bibinfo{year}{1998}).

\bibitem{craswell2007}
\bibinfo{author}{Craswell, N.} \& \bibinfo{author}{Szummer, M.}
\newblock \bibinfo{title}{Random walks on the click graph}.
\newblock In \emph{\bibinfo{booktitle}{Proc. 30th Annual Int. ACM SIGIR
  CRDIR}}, \bibinfo{pages}{239--246} (\bibinfo{organization}{ACM},
  \bibinfo{year}{2007}).

\bibitem{fagin2001}
\bibinfo{author}{Fagin, R.} \emph{et~al.}
\newblock \bibinfo{title}{Random walks with" back buttons"}.
\newblock \emph{\bibinfo{journal}{Ann. Appl. Probab.}}
  \bibinfo{pages}{810--862} (\bibinfo{year}{2001}).

\bibitem{Meiss2010}
\bibinfo{author}{Meiss, M.~R.}, \bibinfo{author}{Gon{\c{c}}alves, B.},
  \bibinfo{author}{Ramasco, J.~J.}, \bibinfo{author}{Flammini, A.} \&
  \bibinfo{author}{Menczer, F.}
\newblock \bibinfo{title}{Agents, bookmarks and clicks: a topical model of web
  navigation}.
\newblock In \emph{\bibinfo{booktitle}{Proc. 21st ACM CHH}},
  \bibinfo{pages}{229--234} (\bibinfo{organization}{ACM},
  \bibinfo{year}{2010}).

\bibitem{Chierichetti2012}
\bibinfo{author}{Chierichetti, F.}, \bibinfo{author}{Kumar, R.},
  \bibinfo{author}{Raghavan, P.} \& \bibinfo{author}{Sarl{\'o}s, T.}
\newblock \bibinfo{title}{Are web users really markovian?}
\newblock In \emph{\bibinfo{booktitle}{Proc. 21st Int. Conf. WWW}},
  \bibinfo{pages}{609--618} (\bibinfo{organization}{ACM},
  \bibinfo{year}{2012}).

\bibitem{Barabasi2005}
\bibinfo{author}{Barab{\'a}si, A.-L.}
\newblock \bibinfo{title}{The origin of bursts and heavy tails in human
  dynamics}.
\newblock \emph{\bibinfo{journal}{Nature}} \textbf{\bibinfo{volume}{435}},
  \bibinfo{pages}{207--211} (\bibinfo{year}{2005}).

\bibitem{Oliveira2005}
\bibinfo{author}{Oliveira, J.~G.} \& \bibinfo{author}{Barab{\'a}si, A.-L.}
\newblock \bibinfo{title}{Human dynamics: Darwin and einstein correspondence
  patterns}.
\newblock \emph{\bibinfo{journal}{Nature}} \textbf{\bibinfo{volume}{437}},
  \bibinfo{pages}{1251--1251} (\bibinfo{year}{2005}).

\bibitem{Dezso2006}
\bibinfo{author}{Dezs{\"o}, Z.} \emph{et~al.}
\newblock \bibinfo{title}{Dynamics of information access on the web}.
\newblock \emph{\bibinfo{journal}{Phys. Rev. E}} \textbf{\bibinfo{volume}{73}},
  \bibinfo{pages}{066132} (\bibinfo{year}{2006}).

\bibitem{Zhou2008}
\bibinfo{author}{Zhou, T.}, \bibinfo{author}{Kiet, H. A.~T.},
  \bibinfo{author}{Kim, B.~J.}, \bibinfo{author}{Wang, B.~H.} \&
  \bibinfo{author}{Holme, P.}
\newblock \bibinfo{title}{Role of activity in human dynamics}.
\newblock \emph{\bibinfo{journal}{Europhys. Lett.}}
  \textbf{\bibinfo{volume}{82}}, \bibinfo{pages}{28002} (\bibinfo{year}{2008}).

\bibitem{Goncalves2008}
\bibinfo{author}{Gon{\c{c}}alves, B.} \& \bibinfo{author}{Ramasco, J.~J.}
\newblock \bibinfo{title}{Human dynamics revealed through web analytics}.
\newblock \emph{\bibinfo{journal}{Phys. Rev. E}} \textbf{\bibinfo{volume}{78}}
  (\bibinfo{year}{2008}).

\bibitem{Wu2010}
\bibinfo{author}{Wu, Y.}, \bibinfo{author}{Zhou, C.}, \bibinfo{author}{Xiao,
  J.}, \bibinfo{author}{Kurths, J.} \& \bibinfo{author}{Schellnhuber, H.~J.}
\newblock \bibinfo{title}{Evidence for a bimodal distribution in human
  communication}.
\newblock \emph{\bibinfo{journal}{Proc. Natl. Acad. Sci. U. S. A.}}
  \textbf{\bibinfo{volume}{107}}, \bibinfo{pages}{18803--18808}
  (\bibinfo{year}{2010}).

\bibitem{Vazquez2006}
\bibinfo{author}{V{\'a}zquez, A.} \emph{et~al.}
\newblock \bibinfo{title}{Modeling bursts and heavy tails in human dynamics}.
\newblock \emph{\bibinfo{journal}{Phys. Rev. E}} \textbf{\bibinfo{volume}{73}},
  \bibinfo{pages}{036127} (\bibinfo{year}{2006}).

\bibitem{Malmgren2008}
\bibinfo{author}{Malmgren, R.~D.}, \bibinfo{author}{Stouffer, D.~B.},
  \bibinfo{author}{Motter, A.~E.} \& \bibinfo{author}{Amaral, L. A.~N.}
\newblock \bibinfo{title}{A poissonian explanation for heavy tails in e-mail
  communication}.
\newblock \emph{\bibinfo{journal}{Proc. Natl. Acad. Sci. U. S. A.}}
  \textbf{\bibinfo{volume}{105}}, \bibinfo{pages}{18153--18158}
  (\bibinfo{year}{2008}).

\bibitem{Malmgren2009}
\bibinfo{author}{Malmgren, R.~D.}, \bibinfo{author}{Stouffer, D.~B.},
  \bibinfo{author}{Campanharo, A. S. L.~O.} \& \bibinfo{author}{Amaral, L.
  A.~N.}
\newblock \bibinfo{title}{On universality in human correspondence activity}.
\newblock \emph{\bibinfo{journal}{Science}} \textbf{\bibinfo{volume}{325}},
  \bibinfo{pages}{1696--1700} (\bibinfo{year}{2009}).

\bibitem{Han2008}
\bibinfo{author}{Han, X.-P.}, \bibinfo{author}{Zhou, T.} \&
  \bibinfo{author}{Wang, B.-H.}
\newblock \bibinfo{title}{Modeling human dynamics with adaptive interest}.
\newblock \emph{\bibinfo{journal}{New J. Phys.}} \textbf{\bibinfo{volume}{10}},
  \bibinfo{pages}{073010} (\bibinfo{year}{2008}).

\bibitem{Vazquez2007}
\bibinfo{author}{V{\'a}zquez, A.}
\newblock \bibinfo{title}{Impact of memory on human dynamics}.
\newblock \emph{\bibinfo{journal}{Physica A}} \textbf{\bibinfo{volume}{373}},
  \bibinfo{pages}{747--752} (\bibinfo{year}{2007}).

\bibitem{Oliveira2009}
\bibinfo{author}{Oliveira, J.~G.} \& \bibinfo{author}{V{\'a}zquez, A.}
\newblock \bibinfo{title}{Impact of interactions on human dynamics}.
\newblock \emph{\bibinfo{journal}{Physica A}} \textbf{\bibinfo{volume}{388}},
  \bibinfo{pages}{187--192} (\bibinfo{year}{2009}).

\bibitem{Min2009}
\bibinfo{author}{Min, B.}, \bibinfo{author}{Goh, K.~I.} \&
  \bibinfo{author}{Kim, I.~M.}
\newblock \bibinfo{title}{Waiting time dynamics of priority-queue networks}.
\newblock \emph{\bibinfo{journal}{Phys. Rev. E}} \textbf{\bibinfo{volume}{79}}
  (\bibinfo{year}{2009}).

\bibitem{Brockmann2006}
\bibinfo{author}{Brockmann, D.}, \bibinfo{author}{Hufnagel, L.} \&
  \bibinfo{author}{Geisel, T.}
\newblock \bibinfo{title}{The scaling laws of human travel}.
\newblock \emph{\bibinfo{journal}{Nature}} \textbf{\bibinfo{volume}{439}},
  \bibinfo{pages}{462--465} (\bibinfo{year}{2006}).

\bibitem{Gonzalez2008}
\bibinfo{author}{Gonzalez, M.~C.}, \bibinfo{author}{Hidalgo, C.~A.} \&
  \bibinfo{author}{Barab{\'a}si, A.-L.}
\newblock \bibinfo{title}{Understanding individual human mobility patterns}.
\newblock \emph{\bibinfo{journal}{Nature}} \textbf{\bibinfo{volume}{453}},
  \bibinfo{pages}{779--782} (\bibinfo{year}{2008}).

\bibitem{Rhee2011}
\bibinfo{author}{Rhee, I.} \emph{et~al.}
\newblock \bibinfo{title}{On the levy-walk nature of human mobility}.
\newblock \emph{\bibinfo{journal}{IEEE/ACM Trans. Net.}}
  \textbf{\bibinfo{volume}{19}}, \bibinfo{pages}{630--643}
  (\bibinfo{year}{2011}).

\bibitem{Song2010}
\bibinfo{author}{Song, C.}, \bibinfo{author}{Koren, T.}, \bibinfo{author}{Wang,
  P.} \& \bibinfo{author}{Barab{\'a}si, A.-L.}
\newblock \bibinfo{title}{Modelling the scaling properties of human mobility}.
\newblock \emph{\bibinfo{journal}{Nat. Phys.}} \textbf{\bibinfo{volume}{6}},
  \bibinfo{pages}{818--823} (\bibinfo{year}{2010}).

\bibitem{Han2011}
\bibinfo{author}{Han, X.-P.}, \bibinfo{author}{Hao, Q.}, \bibinfo{author}{Wang,
  B.-H.} \& \bibinfo{author}{Zhou, T.}
\newblock \bibinfo{title}{Origin of the scaling law in human mobility:
  Hierarchy of traffic systems}.
\newblock \emph{\bibinfo{journal}{Phys. Rev. E}} \textbf{\bibinfo{volume}{83}},
  \bibinfo{pages}{036117} (\bibinfo{year}{2011}).

\bibitem{Yan2011}
\bibinfo{author}{Yan, X.-Y.}, \bibinfo{author}{Han, X.-P.},
  \bibinfo{author}{Zhou, T.} \& \bibinfo{author}{Wang, B.-H.}
\newblock \bibinfo{title}{Exact solution of the gyration radius of an
  individual's trajectory for a simplified human regular mobility model}.
\newblock \emph{\bibinfo{journal}{Chin. Phys. Lett.}}
  \textbf{\bibinfo{volume}{28}}, \bibinfo{pages}{120506}
  (\bibinfo{year}{2011}).

\bibitem{Huberman1998}
\bibinfo{author}{Huberman, B.~A.}, \bibinfo{author}{Pirolli, P. L.~T.},
  \bibinfo{author}{Pitkow, J.~E.} \& \bibinfo{author}{Lukose, R.~M.}
\newblock \bibinfo{title}{Strong regularities in world wide web surfing}.
\newblock \emph{\bibinfo{journal}{Science}} \textbf{\bibinfo{volume}{280}},
  \bibinfo{pages}{95--97} (\bibinfo{year}{1998}).

\bibitem{stehle2010}
\bibinfo{author}{Stehl{\'e}, J.}, \bibinfo{author}{Barrat, A.} \&
  \bibinfo{author}{Bianconi, G.}
\newblock \bibinfo{title}{Dynamical and bursty interactions in social
  networks}.
\newblock \emph{\bibinfo{journal}{Phys. rev. E}} \textbf{\bibinfo{volume}{81}},
  \bibinfo{pages}{035101} (\bibinfo{year}{2010}).

\bibitem{Karsai2012}
\bibinfo{author}{Karsai, M.}, \bibinfo{author}{Kaski, K.},
  \bibinfo{author}{Barab{\'a}si, A.-L.} \& \bibinfo{author}{Kert{\'e}sz, J.}
\newblock \bibinfo{title}{Universal features of correlated bursty behaviour}.
\newblock \emph{\bibinfo{journal}{Scientific Reports}}
  \textbf{\bibinfo{volume}{2}} (\bibinfo{year}{2012}).

\bibitem{Lam2001}
\bibinfo{author}{Lam, W.} \& \bibinfo{author}{Mostafa, J.}
\newblock \bibinfo{title}{Modeling user interest shift using a bayesian
  approach}.
\newblock \emph{\bibinfo{journal}{J. Am. Soc. Inf. Sci. Tech.}}
  \textbf{\bibinfo{volume}{52}}, \bibinfo{pages}{416--429}
  (\bibinfo{year}{2001}).

\bibitem{White2009}
\bibinfo{author}{White, R.~W.}, \bibinfo{author}{Bailey, P.} \&
  \bibinfo{author}{Chen, L.}
\newblock \bibinfo{title}{Predicting user interests from contextual
  information}.
\newblock In \emph{\bibinfo{booktitle}{Proc. 32nd Int. ACM SIGIR CRDIR}},
  \bibinfo{pages}{363--370} (\bibinfo{organization}{ACM},
  \bibinfo{year}{2009}).

\bibitem{Chmiel2009}
\bibinfo{author}{Chmiel, A.}, \bibinfo{author}{Kowalska, K.} \&
  \bibinfo{author}{Ho{\l}yst, J.}
\newblock \bibinfo{title}{Scaling of human behavior during portal browsing}.
\newblock \emph{\bibinfo{journal}{Phys. Rev. E}} \textbf{\bibinfo{volume}{80}},
  \bibinfo{pages}{066122} (\bibinfo{year}{2009}).

\bibitem{Yang2011}
\bibinfo{author}{Yang, S.} \emph{et~al.}
\newblock \bibinfo{title}{Like like alike: joint friendship and interest
  propagation in social networks}.
\newblock In \emph{\bibinfo{booktitle}{Proc. 20th Int. Conf. WWW}},
  \bibinfo{pages}{537--546} (\bibinfo{organization}{ACM},
  \bibinfo{year}{2011}).

\bibitem{Kingman1963}
\bibinfo{author}{Kingman, J. F.~C.}
\newblock \bibinfo{title}{The exponential decay of markov transition
  probabilities}.
\newblock \emph{\bibinfo{journal}{Proc. London Math. Soc.}}
  \textbf{\bibinfo{volume}{3}}, \bibinfo{pages}{337--358}
  (\bibinfo{year}{1963}).

\bibitem{Yamasaki2005}
\bibinfo{author}{Yamasaki, K.}, \bibinfo{author}{Muchnik, L.},
  \bibinfo{author}{Havlin, S.}, \bibinfo{author}{Bunde, A.} \&
  \bibinfo{author}{Stanley, H.~E.}
\newblock \bibinfo{title}{Scaling and memory in volatility return intervals in
  financial markets}.
\newblock \emph{\bibinfo{journal}{Proc. Natl. Acad. Sci. U. S. A.}}
  \textbf{\bibinfo{volume}{102}}, \bibinfo{pages}{9424} (\bibinfo{year}{2005}).

\bibitem{Goh2008}
\bibinfo{author}{Goh, K.~I.} \& \bibinfo{author}{Barab{\'a}si, A.-L.}
\newblock \bibinfo{title}{Burstiness and memory in complex systems}.
\newblock \emph{\bibinfo{journal}{Europhys. Lett.}}
  \textbf{\bibinfo{volume}{81}}, \bibinfo{pages}{48002} (\bibinfo{year}{2008}).

\bibitem{Cai2009}
\bibinfo{author}{Cai, S.~M.}, \bibinfo{author}{Fu, Z.~Q.},
  \bibinfo{author}{Zhou, T.}, \bibinfo{author}{Gu, J.} \&
  \bibinfo{author}{Zhou, P.~L.}
\newblock \bibinfo{title}{Scaling and memory in recurrence intervals of
  internet traffic}.
\newblock \emph{\bibinfo{journal}{Europhys. Lett.}}
  \textbf{\bibinfo{volume}{87}}, \bibinfo{pages}{68001} (\bibinfo{year}{2009}).

\bibitem{Szell2012}
\bibinfo{author}{Szell, M.}, \bibinfo{author}{Sinatra, R.},
  \bibinfo{author}{Petri, G.}, \bibinfo{author}{Thurner, S.} \&
  \bibinfo{author}{Latora, V.}
\newblock \bibinfo{title}{Understanding mobility in a social petri dish}.
\newblock \emph{\bibinfo{journal}{Scientific Reports}}
  \textbf{\bibinfo{volume}{2}} (\bibinfo{year}{2012}).

\bibitem{Busemeyer1993}
\bibinfo{author}{Busemeyer, J.~R.} \& \bibinfo{author}{Townsend, J.~T.}
\newblock \bibinfo{title}{Decision field theory}.
\newblock \emph{\bibinfo{journal}{Psychological Review}}
  \textbf{\bibinfo{volume}{100}}, \bibinfo{pages}{432--59}
  (\bibinfo{year}{1993}).

\bibitem{salganik2006}
\bibinfo{author}{Salganik, M.~J.}, \bibinfo{author}{Dodds, P.} \&
  \bibinfo{author}{Watts, D.~J.}
\newblock \bibinfo{title}{Experimental study of inequality and unpredictability
  in an artificial cultural market}.
\newblock \emph{\bibinfo{journal}{Science}} \textbf{\bibinfo{volume}{311}},
  \bibinfo{pages}{854--856} (\bibinfo{year}{2006}).

\bibitem{Banavar1999}
\bibinfo{author}{Banavar, J.~R.}, \bibinfo{author}{Maritan, A.} \&
  \bibinfo{author}{Rinaldo, A.}
\newblock \bibinfo{title}{Size and form in efficient transportation networks}.
\newblock \emph{\bibinfo{journal}{Nature}} \textbf{\bibinfo{volume}{399}},
  \bibinfo{pages}{130--132} (\bibinfo{year}{1999}).

\bibitem{Rosvall2008}
\bibinfo{author}{Rosvall, M.} \& \bibinfo{author}{Bergstrom, C.~T.}
\newblock \bibinfo{title}{Maps of random walks on complex networks reveal
  community structure}.
\newblock \emph{\bibinfo{journal}{Proc. Natl. Acad. Sci. U. S. A.}}
  \textbf{\bibinfo{volume}{105}}, \bibinfo{pages}{1118--1123}
  (\bibinfo{year}{2008}).

\bibitem{Song2010a}
\bibinfo{author}{Song, C.}, \bibinfo{author}{Qu, Z.}, \bibinfo{author}{Blumm,
  N.} \& \bibinfo{author}{Barab{\'a}si, A.-L.}
\newblock \bibinfo{title}{Limits of predictability in human mobility}.
\newblock \emph{\bibinfo{journal}{Science}} \textbf{\bibinfo{volume}{327}},
  \bibinfo{pages}{1018--1021} (\bibinfo{year}{2010}).

\bibitem{Bagrow2012}
\bibinfo{author}{Bagrow, J.~P.} \& \bibinfo{author}{Lin, Y.-R.}
\newblock \bibinfo{title}{Mesoscopic structure and social aspects of human
  mobility}.
\newblock \emph{\bibinfo{journal}{PLoS ONE}} \textbf{\bibinfo{volume}{7}},
  \bibinfo{pages}{e37676} (\bibinfo{year}{2012}).

\bibitem{Barabasi1999}
\bibinfo{author}{Barab{\'a}si, A.-L.} \& \bibinfo{author}{Albert, R.}
\newblock \bibinfo{title}{Emergence of scaling in random networks}.
\newblock \emph{\bibinfo{journal}{Science}} \textbf{\bibinfo{volume}{286}},
  \bibinfo{pages}{509--512} (\bibinfo{year}{1999}).

\bibitem{Gonalves2009}
\bibinfo{author}{Gon{\c{c}}alves, B.}, \bibinfo{author}{Meiss, M.~R.},
  \bibinfo{author}{Ramasco, J.~J.}, \bibinfo{author}{Flammini, A.} \&
  \bibinfo{author}{Menczer, F.}
\newblock \bibinfo{title}{Remembering what we like: Toward an agent-based model
  of web traffic}.
\newblock In \emph{\bibinfo{booktitle}{The 2nd ACM Int. Conf. WSDM}}
  (\bibinfo{organization}{ACM}, \bibinfo{year}{2009}).

\bibitem{Sorribes2011}
\bibinfo{author}{Sorribes, A.}, \bibinfo{author}{Armendariz, B.~G.},
  \bibinfo{author}{Lopez-Pigozzi, D.}, \bibinfo{author}{Murga, C.} \&
  \bibinfo{author}{de~Polavieja, G.~G.}
\newblock \bibinfo{title}{The origin of behavioral bursts in decision-making
  circuitry}.
\newblock \emph{\bibinfo{journal}{PLoS Comp. Bio.}}
  \textbf{\bibinfo{volume}{7}}, \bibinfo{pages}{e1002075}
  (\bibinfo{year}{2011}).

\bibitem{Antal2005}
\bibinfo{author}{Antal, T.} \& \bibinfo{author}{Redner, S.}
\newblock \bibinfo{title}{The excited random walk in one dimension}.
\newblock \emph{\bibinfo{journal}{J. Phys. A}} \textbf{\bibinfo{volume}{38}},
  \bibinfo{pages}{2555} (\bibinfo{year}{2005}).

\bibitem{Vespignani2009}
\bibinfo{author}{Vespignani, A.}
\newblock \bibinfo{title}{Predicting the behavior of techno-social systems}.
\newblock \emph{\bibinfo{journal}{Science}} \textbf{\bibinfo{volume}{325}},
  \bibinfo{pages}{425--428} (\bibinfo{year}{2009}).

\bibitem{Balcan2011}
\bibinfo{author}{Balcan, D.} \& \bibinfo{author}{Vespignani, A.}
\newblock \bibinfo{title}{Phase transitions in contagion processes mediated by
  recurrent mobility patterns}.
\newblock \emph{\bibinfo{journal}{Nat. Phys.}} \textbf{\bibinfo{volume}{7}},
  \bibinfo{pages}{581--586} (\bibinfo{year}{2011}).

\bibitem{Zhao2012}
\bibinfo{author}{Zhao, Z.-D.}, \bibinfo{author}{Liu, Y.} \&
  \bibinfo{author}{Tang, M.}
\newblock \bibinfo{title}{Epidemic variability in hierarchical geographical
  networks with human activity patterns}.
\newblock \emph{\bibinfo{journal}{Chaos}} \textbf{\bibinfo{volume}{22}},
  \bibinfo{pages}{023150} (\bibinfo{year}{2012}).

\bibitem{Spiliopoulou2003}
\bibinfo{author}{Spiliopoulou, M.}, \bibinfo{author}{Mobasher, B.},
  \bibinfo{author}{Berendt, B.} \& \bibinfo{author}{Nakagawa, M.}
\newblock \bibinfo{title}{A framework for the evaluation of session
  reconstruction heuristics in web-usage analysis}.
\newblock \emph{\bibinfo{journal}{INFORMS J. Comp.}}
  \textbf{\bibinfo{volume}{15}}, \bibinfo{pages}{171--190}
  (\bibinfo{year}{2003}).

\bibitem{Borges2007}
\bibinfo{author}{Borges, J.} \& \bibinfo{author}{Levene, M.}
\newblock \bibinfo{title}{Evaluating variable-length markov chain models for
  analysis of user web navigation sessions}.
\newblock \emph{\bibinfo{journal}{IEEE Trans. Knowl. Data En.}}
  \textbf{\bibinfo{volume}{19}}, \bibinfo{pages}{441--452}
  (\bibinfo{year}{2007}).

\bibitem{Meiss2009}
\bibinfo{author}{Meiss, M.}, \bibinfo{author}{Duncan, J.},
  \bibinfo{author}{Gon{\c{c}}alves, B.}, \bibinfo{author}{Ramasco, J.~J.} \&
  \bibinfo{author}{Menczer, F.}
\newblock \bibinfo{title}{What's in a session: tracking individual behavior on
  the web}.
\newblock In \emph{\bibinfo{booktitle}{Proc. 20th ACM CHH}},
  \bibinfo{pages}{173--182} (\bibinfo{organization}{ACM},
  \bibinfo{year}{2009}).

\bibitem{Fortunato2006}
\bibinfo{author}{Fortunato, S.}, \bibinfo{author}{Flammini, A.},
  \bibinfo{author}{Menczer, F.} \& \bibinfo{author}{Vespignani, A.}
\newblock \bibinfo{title}{Topical interests and the mitigation of search engine
  bias}.
\newblock \emph{\bibinfo{journal}{Proc. Natl. Acad. Sci. U. S. A.}}
  \textbf{\bibinfo{volume}{103}}, \bibinfo{pages}{12684--12689}
  (\bibinfo{year}{2006}).

\bibitem{Zhou2012}
\bibinfo{author}{Zhou, T.}, \bibinfo{author}{Zhao, Z.-D.},
  \bibinfo{author}{Yang, Z.} \& \bibinfo{author}{Zhou, C.}
\newblock \bibinfo{title}{Relative clock verifies endogenous bursts of human
  dynamics}.
\newblock \emph{\bibinfo{journal}{Europhys. Lett.}}
  \textbf{\bibinfo{volume}{97}}, \bibinfo{pages}{18006} (\bibinfo{year}{2012}).

\bibitem{Clauset2009}
\bibinfo{author}{Clauset, A.}, \bibinfo{author}{Shalizi, C.} \&
  \bibinfo{author}{Newman, M. E.~J.}
\newblock \bibinfo{title}{Power-law distributions in empirical data}.
\newblock \emph{\bibinfo{journal}{SIAM Review}} \textbf{\bibinfo{volume}{51}},
  \bibinfo{pages}{661--703} (\bibinfo{year}{2009}).

\end{thebibliography}

\newpage

\end{document}